# Structure and superconductivity of LiFeAs

Michael J. Pitcher,[†] Dinah R. Parker,[†] Paul Adamson,[†] Sebastian J. C. Herkelrath,[†] Andrew T. Boothroyd[‡] and Simon J. Clarke*[†]

*Department of Chemistry, University of Oxford, Inorganic Chemistry Laboratory, South Parks Road, Oxford, OX1 3QR, UK.; Department of Physics, University of Oxford, Clarendon Laboratory, Parks Road, Oxford, OX1 3PU, UK.*

**Lithium iron arsenide phases with compositions close to LiFeAs exhibit superconductivity at temperatures at least as high as 16 K demonstrating that superconducting [FeAs]⁻ anionic layers with the anti-PbO structure type occur in at least three different structure types and with a wide range of As−Fe−As bond angles.**

High temperature superconductivity has recently been reported in several compounds containing FeAs anti-PbO-type (i.e. antifluorite-type) layers. LaOFeAs with the ZrSiCuAs structure type was found to superconduct when doped with electrons through the substitution of about 10-20 % of the oxide ions by fluoride.[1] This resulted in superconductivity below about 26 K, a critical temperature, $T_c$, only exceeded by layered cuprate superconductors, some fullerides and $MgB_2$. Enhancement of $T_c$ to 43 K was achieved in this system at high pressure.[2] Substitution of lanthanum in $LaO_{1-x}F_xFeAs$ by heavier and smaller lanthanides[3–5] results in higher $T_c$s, and it has been shown that electron doping by the introduction of oxygen deficiency leads to $T_c$s of up to 55 K in $SmO_{1-x}FeAs$.[6] The high $T_c$s and critical fields exhibited by these superconductors and their proximity to magnetically ordered phases[7] suggest that they are unconventional superconductors whose properties cannot be described within the framework of existing models of superconductivity such as the BCS theory.

$BaFe_2As_2$ with the common $ThCr_2Si_2$ structure type was found to exhibit similar magnetic and structural behaviour to LaOFeAs.[8] The two compounds are formally isoelectronic as far as their FeAs layers are concerned. Reducing the electron count of $BaFe_2As_2$ to form $Ba_{1-x}K_xFe_2As_2$ produced superconductivity below 38 K[9] and similar doping of $SrFe_2As_2$ also produces superconductivity below 38 K.[10] Superconductivity has been induced at high pressure in stoichiometric $AFe_2As_2$ ($A$ = Ca, Sr, Ba).[11] Here we describe superconductivity in a sample of almost stoichiometric LiFeAs which represents a third structure type in which superconducting anti-PbO-type FeAs layers occur.

Investigations by Juza and Langer[12] indicated that the composition LiFeAs could not be obtained but that both Fe-rich and Li-rich compositions could be prepared; for metal:arsenic ratios of 2:1 single phase compositions were reported between the limits $Li_{1.1}FeAs$ and $Li_{0.96}Fe_{1.14}As$. We initially synthesised a sample (Sample 1) with composition "LiFeAs" by the reaction at 800 °C between stoichiometric quantities of elemental lithium and FeAs (previously prepared from the elements) in a tantalum tube sealed by welding under 1 atm of argon gas. According to powder X-ray diffraction (PXRD) measurements, this black sample was composed of a highly air-sensitive ternary $Li_xFeAs$ phase (84(1) mol %) with lattice parameters $a$ = 3.775(1) Å and $c$ = 6.358(2) Å, and 16(1) mol % FeAs. Magnetometry measurements (Quantum Design MPMS5) showed a sharp superconducting transition with $T_c$ = 16 K (Figure 1) and the shielding volume fraction was estimated to be close to 100 %. Subsequently we attempted the synthesis of 3 g of material of composition $Li_{1.1}FeAs$ using a method similar to that described by Juza and Langer[12] in which $Li_3As$, Fe and FeAs were ground together in the appropriate stoichiometric ratio, pressed into a pellet and heated at 800 °C for 48 hours in an alumina crucible sealed inside an evacuated silica ampoule. Some attack of the silica tube, presumably by Li was evident, but the black product (Sample 2) appeared single phase according to PXRD measurements with lattice parameters very similar to those of Sample 1: $a$ = 3.774(1) Å and $c$ = 6.354(2) Å. Sample 2 exhibited superconductivity below 10 K, although the shielding volume fraction was much smaller than that of Sample 1 and the superconducting transition was slightly less sharp (Figure 1).

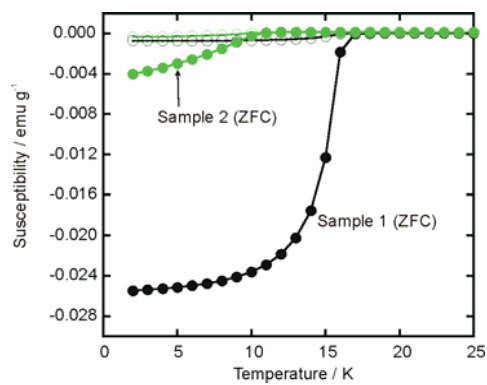

***Figure 1.*** Magnetic susceptibilities of Samples 1 (black symbols) and 2 (green symbols) measured in applied fields of 50 Oe after cooling in zero applied field (ZFC; closed symbols) and after cooling in the measuring field (open symbols).

Sample 2 was measured at 298 K and at 6.5 K on the high resolution time of flight powder neutron diffractometer HRPD at the ISIS facility, UK (Figure 2). Analysis of the data (0.4 < $d$ < 7 Å) using the GSAS suite showed unequivocally that lithium ions were located in square pyramidal positions 5-coordinated by As as proposed by Juza and Langer.[12] The structure of the compound is quite different from that obtained for LiMnAs[13] in which Fe and Li occupy alternate layers of tetrahedral sites in an approximately cubic close packed array of As. The location of the lithium ions in LiFeAs is similar to that of half of the Fe ions in $Fe_2As$ ($Cu_2Sb$ type). The refined crystal structure is compared with the crystal structures of $Fe_2As$ and LaOFeAs in Figure 3.

Refinement of the structure at 6.5 K indicated no structural distortion, in contrast to the behaviour of non-superconducting LaOFeAs and $AFe_2As_2$ ($A$ = Sr, Ba). Nor was any magnetic Bragg scattering evident.

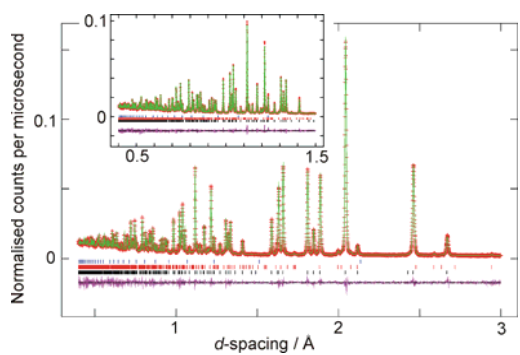

**Figure 2.** Results of Rietveld refinement at room temperature against HRPD data (90 degree bank and 168 degree bank inset). The data (red points), fit (green line) and difference (lower purple line) are shown. Tick marks indicate reflections for LiFeAs (lowest), FeAs (1.5 mol %; middle) and the vanadium container (uppermost). Space group $P4/nmm$ (No. 129) $a$ = 3.77531(1) Å, $c$ = 6.35326(2) Å, $Z$ = 2; $wR_p$ = 0.0389, $\chi^2$ = 7.035.

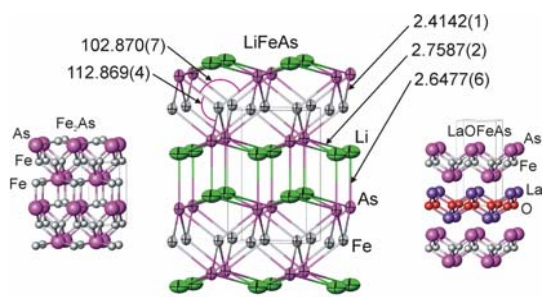

**Figure 3.** Comparison of the crystal structures of $Fe_2As$, LiFeAs, and LaOFeAs. In LiFeAs lithium ions fully occupy one of the sites which is occupied by Fe in $Fe_2As$ and Li is five-coordinate (square-based pyramid) by As. The interatomic distances (Å) and bond angles (degrees) for LiFeAs are those obtained from refinement against room temperature PND data. 99 % anisotropic displacement ellipsoids are shown.

The refinements at room temperature and 6.5 K both suggested an apparent deficiency of 4 % on the Li site and 2-3 % on the As site. The refined composition was $Li_{0.963(4)}FeAs_{0.972(1)}$ at 298 K and $Li_{0.950(4)}FeAs_{0.976(1)}$ at 6.5 K where the uncertainties in the fractional occupancies indicate the statistical precision inherent in the refinement.[14] The agreement in the refined compositions obtained at two temperatures suggests that the data are of sufficient quality to enable fractional site occupancies and anisotropic displacement parameters to be refined independently. Because of the difference in scattering lengths of Fe (9.54 fm) and Li (−1.90 fm) an apparent 4 % deficiency on the Li site could result from accommodating 0.7 % Fe atoms and 99.3 % Li atoms on that site, which is chemically plausible. Furthermore, from a combination of laboratory PXRD and the PND measurements we cannot rule out a small population of the site (¾, ¼, ½) by a null-scattering mixture of up to 1% Fe and 5 % Li. Therefore the composition of Sample 2 is close to LiFeAs but with an uncertainty of about 5 % on the Li content and a 3(1) % As deficiency indicated by the PND refinement. According to the PND refinement the Sample 2 contained 1.5 mol % of FeAs impurity.

The published structure of LiFeAs[12] is approximate because of the almost negligible X-ray scattering amplitude of lithium. Our analysis of PND data enables accurate location of the lithium ions which are in sites surrounded by a square pyramid of As atoms (Li−As distances are 2.6477(6) Å (× 1) and 2.7587(2) Å (× 4) as shown in Figure 3). Relative to the Fe atoms, the Li atoms in LiFeAs are in a similar location to the La atoms in LaOFeAs. The structure contains FeAs anti-PbO type layers in common with other recently reported superconductors. Our refinement of the structure of Sample 2 produced an Fe−As distance of 2.4142(1) Å (× 4) at room temperature which is similar to the distances of 2.407(2) Å in LaOFeAs and 2.388(3) Å in $SrFe_2As_2$.[15] However the edge-sharing $FeAs_4$ tetrahedra in LiFeAs are compressed in the basal plane relative to the tetrahedra in LaOFeAs and $SrFe_2As_2$: while the As−Fe−As angles in LaOFeAs[7] are 113.7° (× 2) and 107.41° (× 4) (i.e. compressed perpendicular to the basal plane), and the $FeAs_4$ tetrahedra in $SrFe_2As_2$[15] are almost regular (angles of 110.5° (× 2) and 108.9° (× 4)), the corresponding angles in LiFeAs are 102.870(7)° (× 2) and 112.869(4)° (× 4). Indeed LiFeAs has the smallest basal lattice parameter of all the superconducting iron arsenide phases so far reported and the *occurrence* of superconductivity in compounds containing such layers does not seem highly sensitive to the As−Fe−As bond angles, although the coordination environment is expected to influence the value of $T_c$.

Our results suggests that other compounds containing FeAs anti-PbO-type layers with Fe in a formal oxidation state close to +2 should be synthesised and their electronic and magnetic properties investigated. Further detailed structural analysis of other phases in the Li-Fe-As system is required and correlation of composition and structure with $T_c$ are in progress.

During the preparation of this manuscript we became aware of the work of Jin and co-workers[16] which reported superconductivity below 18 K in samples containing $Li_xFeAs$ phases and FeAs as an impurity phase synthesised using high pressure techniques.

**Acknowledgement.** We thank the UK EPSRC for funding under grant EP/E025447 and for access to the ISIS facility. We are grateful to Dr R. M. Ibberson, Dr R. I. Smith and Dr K. Knight (ISIS facility) for performing the neutron data collection.

| Instrument | HRPD | |
|---|---|---|
| Temperature / K | 298 | 6.5 |
| Space group | P4/nmm (No. 129) | |
| a / Å | 3.77543(3) | 3.76982(4) |
| c / Å | 6.35345(6) | 6.30693(7) |
| V / Å$^3$ | 90.561(1) | 89.631(2) |
| R$_{wp}$ | 0.0389 | 0.0304 |
| $\chi^2$ | 7.035 | 8.115 |

**Table 1.** Summary of refined parameters for Sample 2 against powder neutron diffraction data

| Atom | Site | x | y | z | U$_{equiv}$ / Å$^2$ × 100 | Refined fractional occupancy |
|---|---|---|---|---|---|---|
| Fe | 2a | 0.75 | 0.25 | 0 | 0.66(1) | 1 |
|  |  |  |  |  | 0.18(1) | 1 |
| Li | 2c | 0.25 | 0.25 | 0.6538(1) | 1.61(5) | 0.963(4) |
|  |  |  |  | 0.6554(2) | 0.77(6) | 0.950(4) |
| As | 2c | 0.25 | 0.25 | 0.23685(4) | 0.62(2) | 0.972(1) |
|  |  |  |  | 0.23646(5) | 0.12(1) | 0.976(1) |

**Table 2.** Refined atomic parameters at 298 K (upper) and 6.5 K (lower)

| Atom | U$_{11}$ = U$_{22}$ / Å$^2$ × 100 | U$_{33}$ / Å$^2$ × 100 |
|---|---|---|
| Fe | 0.580(8) | 0.82(1) |
|  | 0.184(7) | 0.177(9) |
| Li | 1.95(4) | 0.94(6) |
|  | 0.90(4) | 0.50(6) |
| As | 0.535(9) | 0.80(2) |
|  | 0.089(9) | 0.18(1) |

**Table 3**. Refined anisotropic displacement ellipsoids at 298 K (upper) and 6.5 K (lower)